\documentclass[pdftex,twocolumn,epjc3,preprintnumbers,amsmath,amssymb]{svjour3}          

\RequirePackage[T1]{fontenc}

\smartqed  

\RequirePackage{graphicx}
\RequirePackage{mathptmx}   
\RequirePackage{flushend}
\RequirePackage[numbers,sort&compress]{natbib}
\RequirePackage[colorlinks,citecolor=blue,urlcolor=blue,linkcolor=blue]{hyperref}

\journalname{Eur. Phys. J. C}
\usepackage{amsmath}
\usepackage{amssymb}
\usepackage[normalem]{ulem}
\usepackage{bm}
\usepackage{color}
\usepackage{graphicx}
\usepackage[colorlinks,citecolor=blue,urlcolor=blue,linkcolor=blue]{hyperref}
\usepackage{multirow}
\usepackage{subfigure}
\usepackage{booktabs}
\usepackage{mathrsfs}
\usepackage{balance}

\newcommand{\f}{\frac}
\newcommand{\lt}{\left}

\newcommand{\n}{\nonumber}

\newcommand{\rt}{\right}
\newcommand{\dd}{{\rm d}}
\newcommand{\bt}{\beta}
\newcommand{\dt}{\delta}

\newcommand{\sg}{\sigma}

\newcommand{\pb}{{\rm PBH}}

\begin{document}
\title{Constraining primordial black holes as dark matter using AMS-02 data} 


\author{Bing-Yu Su\thanksref{addr1,addr2} 
\and 
Xu Pan\thanksref{addr1,addr2,e1} 
\and 
Guan-Sen Wang\thanksref{addr1,addr2} 
\and 
Lei Zu\thanksref{addr1,addr2,e2}
\and 
Yupeng Yang\thanksref{addr3,addr4,e3}
\and 
Lei Feng\thanksref{addr1,addr2,addr4,e4}} 

\thankstext{e1}{e-mail: xupan@pmo.ac.cn (corresponding author)}
\thankstext{e2}{e-mail: zulei@pmo.ac.cn (corresponding author)}
\thankstext{e3}{e-mail: ypyang@qfnu.edu.cn (corresponding author)}
\thankstext{e4}{e-mail: fenglei@pmo.ac.cn (corresponding author)}

\institute{Key Laboratory of Dark Matter and Space Astronomy, Purple Mountain Observatory, Chinese Academy of Sciences, Nanjing 210023, China\label{addr1} 
\and 
School of Astronomy and Space Science, University of Science and Technology of China, Hefei, Anhui 230026, China\label{addr2} 
\and 
School of Physics and Physical Engineering, Qufu Normal University, Qufu, Shandong, 273165, China\label{addr3} 
\and 
Joint Center for Particle, Nuclear Physics and Cosmology,  Nanjing University--Purple Mountain Observatory,  Nanjing  210093, China\label{addr4} }

\date{Received: date / Accepted: date}

\maketitle

\begin{abstract}
Primordial black holes (PBHs) are the plausible candidates for the cosmological dark matter. Theoretically, PBHs with masses $M_{\rm PBH}$ in the range of $4\times10^{14}\sim 10^{17}\,{\rm g}$ can emit sub-GeV electrons and positrons through Hawking radiation. Some of these particles could undergo diffusive reacceleration during propagation in the Milky Way, potentially reaching energies up to the GeV level observed by AMS-02. In this work, we utilize AMS-02 data to constrain the PBH abundance $f_{\rm PBH}$ by employing the reacceleration mechanism. Under the assumption of a monochromatic PBH mass distribution, our findings reveal that the limit is stricter than that derived from Voyager 1 data. This difference is particularly pronounced when $M_{\rm PBH}\lesssim10^{15}\,{\rm g}$, exceeding an order of magnitude. The constraints are even more robust in a more realistic scenario involving a log-normal mass distribution of PBHs. Moreover, we explore the impact of varying propagation parameters and solar modulation potential within reasonable ranges, and find that such variations have minimal effects on the final results.

\end{abstract}

\section{Introduction} \label{sec:intro}

Dark matter (DM) accounts for 26\% of the current universe, and yet its precise nature remains elusive. Various plausible candidates are proposed, such as weakly interacting massive particles, axions, fuzzy DM, etc \cite{Peter:2012rz, Bertone:2016nfn, Arun:2017uaw}. Primordial black holes (PBHs) can also partially or fully constitute DM depicted by the PBH abundance $f_\pb$, which have been extensively investigated \cite{Carr:2016drx, Carr:2020xqk, Carr:2020gox, Green:2020jor, Frampton:2010sw, Belotsky:2014kca, Heydari:2021gea, Heydari:2021qsr, Wang:2021kbh, Liu:2021qky, Zhao:2023xnh, Heydari:2023xts, Heydari:2023rmq, Sharma:2024whg}. In general, PBHs can be considered viable DM candidates only when $f_{\rm PBH}\gtrsim 0.1$. 

PBHs are formed from the collapse of large density perturbations in the early universe with a broad mass range of about $10^{-5}\sim 10^{55}\,{\rm g}$ \cite{Zeldovich:1967lct, Hawking:1971ei, Carr:1974nx}. Small PBHs are expected to experience mass loss through the emission of quantum fields with a quasi-thermal spectrum, which is known as Hawking radiation \cite{Hawking:1974rv, Hawking:1975vcx}. The lost energy will be emitted into the environment in the form of particles such as photons, electrons/positrons and neutrinos, which is detectable through the cosmological and astrophysical observations \cite{Laha:2019ssq, Arbey:2019vqx, Ballesteros:2019exr, laha2020n, Okele:1980kwj, okeke1980primary, MacGibbon:1991vc, Boudaud:2018hqb,Dasgupta:2019cae, Wang:2020uvi, Calabrese:2021zfq, Bernal:2022swt, Liu:2023cqs, Huang:2024xap}. Theoretically, black holes with lower mass tend to evaporate more rapidly. Previous researches have investigated the PBHs with masses less than $4\times10^{14}\,{\rm g}$, exploring their cosmological effects at early universe as an additional energy source \cite{Capanema:2021hnm, Carr:2009jm, Auffinger:2022khh}. These investigations include implications for phenomena such as the big bang nucleosynthesis \cite{Kohri:1999ex, Keith:2020jww} and the cosmic microwave background \cite{Clark:2016nst, Acharya:2020jbv}. Furthermore, for the surviving PBHs with masses $4\times10^{14}\,{\rm g}\lesssim M_\pb\lesssim 10^{17}\,{\rm g}$, Hawking radiation remains pronounced within our local universe and is potentially detectable through the emission of its final particles \cite{Clark:2018ghm, Chan:2020zry, Mittal:2021egv, Yang:2022puh}.

Our investigation specifically focuses on the electrons and positrons emitted from PBHs, aiming to impose constraints on PBH abundance $f_\pb$ at the mass range $4\times10^{14}\,{\rm g}\lesssim M_\pb\lesssim 10^{17}\,{\rm g}$. In contrast to photons and neutrinos, the energy of these cosmic ray (CR) electrons/positrons is subject to environmental influences. Previous studies have primarily concentrated on energy loss resulting from interactions with the interstellar medium and radiation during propagation. However, the impact of random shocks in interstellar space, capable of reaccelerating low-energy CR particles, has gained attention. This reacceleration mechanism has been extensively discussed and has demonstrated its ability to self-consistently fit both the proton and Boron-to-Carbon ratio \cite{Yuan:2017ozr, Yuan:2018lmc, Zu:2017dzm, Zu:2021odn}. Furthermore, this investigation presents a novel opportunity for exploring electrons and positrons emitted from PBHs with energies even lower than the detectability threshold of the AMS-02 experiment, i.e., the lower energy CRs undergo a process of reacceleration during propagation, expanding the observable energy range to a higher value.

In our study, we employ a meticulous calculation method incorporating the positron fraction data \cite{AMS:2013fma} and the combined electron and positron data \cite{AMS:2014gdf} from AMS-02  to impose limits on PBH abundance. This allows us to constrain $f_\pb$ with increased precision compared to using the Voyager 1 data of all-electrons \cite{Stone:2013zlg, Cummings:2016pdr}, as the AMS-02 data offers higher accuracy. Our approach involves leveraging more accurate numerical calculation tools and incorporating updated propagation parameters \cite{Fermi-LAT:2012pls}. This refined methodology aims to provide a more comprehensive and precise assessment of the $f_\pb$ constraints based on the latest observational data. We employ the LikeDM code \cite{Huang:2016pxg} to calculate the flux of electrons and positrons produced by PBHs, and get new limits on PBH abundance using the AMS-02 data.

This paper is organized as follows. In Sec. \ref{sec:pbh}, we briefly review the production of positrons and electrons through Hawking radiation. In Sec. \ref{sec:prop}, the propagation process of positrons and electrons is described. Sec. \ref{sec:res} presents the results, followed by a conclusion in Sec. \ref{sec:con}. 

\section{Electrons and positrons from evaporating PBHs}
\label{sec:pbh}

Black holes, including PBHs, can emit particles through a phenomenon known as Hawking radiation. The emitted radiation is characterized as thermal radiation, exhibiting a temperature that is inversely proportional to the mass of the black hole. 
The non-rotating PBH temperature $T_{\rm H}$ is described as \cite{Hawking:1974rv, Hawking:1975vcx}
\begin{align}
k_{\rm B}T_{\rm H}=\f{\hslash c^3}{8\pi GM_{\pb}}\approx1.06\lt(\f{10^{13}\,{\rm g}}{M_{\pb}}\rt)\,{\rm GeV}, \n
\end{align}
where $M_\pb$ is the PBH mass. The corresponding primary emission spectrum of emitted electrons and positrons can be expressed as~\cite{MacGibbon:1990zk}
\begin{align}
\f{\dd^2 N_e}{\dd E\,\dd t}=\f{\Gamma_e}{2\pi}\lt[\exp{\lt(\f{E}{k_{\rm B}T_{\rm H}}\rt)}+1\rt]^{-1}, \n
\end{align}
where $\Gamma_e$ is the electron absorption probability, approximately modeled as
\begin{align}
\Gamma_e=\f{27G^2M_\pb^2E^2}{\hslash^3c^6}\n
\end{align}
at high energies 
for $G M_\pb E/(\hslash c^3)\gg 1$ 
\cite{MacGibbon:1990zk}. In this work, we only consider the non-rotating PBHs with primary electron/positron emission. For one thing, the spin of PBHs introduces a slight modification to the injected energy spectrum~\cite{Dasgupta:2019cae}, and this alteration tends to smooth out as the propagation. For another, the contribution of secondary electrons and positrons resulting from the decay of unstable particles is found to be negligible, constituting only 1\% of the total flux. This calculation has been performed using the publicly available code BlackHawk~\cite{Arbey:2019mbc, Arbey:2021mbl}.

In general, the Hawking radiation of PBHs provides an electron/positron source, described as 
\begin{align}
Q(E,r)=\f{\rho_\pb(r)}{\rho_\odot}\int_{\forall M_\pb}\dd M_\pb\,\f{g(M_\pb)}{M_\pb}\f{\dd^2 N_e}{\dd E\,\dd t}, \label{Qall}
\end{align}
where $g(M_\pb)$ represents the mass distribution of PBHs normalized to $\rho_\odot$, and $\rho_\pb(r)$ denotes the total PBH mass density as a function of the distance from the Milky Way center $r$. It is assumed that $\rho_\pb(r)$ traces the DM density $\rho_{\rm DM}(r)$, and can be expressed as $\rho_\pb(r)=f_\pb\rho_{\rm DM}(r)$, with $f_\pb$ being the PBH abundance. Under the hypothesis of a common mass for all PBHs (i.e., a monochromatic mass distribution), Eq. (\ref{Qall}) is reduced to
\begin{align}
Q(E,r)=\f{\rho_\pb(r)}{M_\pb}\f{\dd^2 N_e}{\dd E\,\dd t}. \n
\end{align}
In this work, we also concentrate on log-normal distribution, defined as \cite{Krishnamoorthy}
\begin{align}
g(M_\pb)=\f{\rho_\odot}{\sqrt{2\pi}\sigma M_\pb}\exp \lt[-\f{\ln^2(M_\pb/\mu)}{2\sg^2}\rt], \label{log}
\end{align}
where $\mu$ is the mass for which the density is maximal, and $\sg$ is the width. 


\section{The propagation of evaporated electrons and positrons}
\label{sec:prop}

The Hawking radiation from PBHs serves as an additional source of electrons and positrons, and these particles undergo diffusive propagation within the Milky Way as part of CRs. We first calculate the primary emission spectra by using the BlackHawk \cite{Arbey:2019mbc, Arbey:2021mbl}. Subsequently, we integrate this information as the CR source into the propagation calculation code. Numerical tools, such as GALPROP \cite{Strong:1998pw} and DRAGON \cite{Evoli:2008dv}, have been developed for CR propagation calculations. In this work, we employ the public code LikeDM \cite{Huang:2016pxg} to calculate the propagation process.
For a specified source distribution, the LikeDM code employs the Green's function method, relying on numerical tables obtained from GALPROP, to calculate the propagation process. This approach has been verified to provide a good approximation to the GALPROP output while being significantly more efficient.

The propagation is assumed to occur within a diffusion reacceleration framework, with the determination of propagation parameters relying on the boron-to-carbon ratio data and the diffuse $\gamma$-ray emission observed by Fermi-LAT~\cite{Fermi-LAT:2012edv}. It is worth noting that the chosen parameters have been updated compared to Ref. \cite{Boudaud:2018hqb}. The main propagation parameters are shown in Tab. \ref{tab:1}, including the diffusion coefficient $D_{xx}$, the characteristic halo height $z_{\rm h}$ and the Alfvenic speed $v_{\rm A}$ describing the reacceleration effect. Additionally, we apply the simple force-field approximation \cite{Gleeson:1968zza} with a broad range of modulation potential to describe the solar modulation. The PBH density is modeled using the Navarro--Frenk--White (NFW) profile \cite{Navarro:1996gj}
\begin{align}
\rho_{\rm NFW}=\rho_{\rm s}\f{r_{\rm s}}{r\lt(1+\f{r}{r_{\rm s}}\rt)^2}, \n
\end{align}
where $r_{\rm s}=20\,{\rm kpc}$ and $\rho_{\rm s}=0.26\,{\rm GeV}\,{\rm cm}^{-3}$ represent the scale radius and scale density \cite{Bertone:2008xr}, respectively. 
\begin{table}
\renewcommand\arraystretch{1.3}
\centering
\begin{tabular}
{ccccc}
\toprule
 & $D_0$ & $z_{\rm h}$ & $v_{\rm A}$ & $\dt$ \\
 & (28 ${\rm cm}^2\,{\rm s}^{-1}$) & (kpc) & (km s$^{-1}$) & \\
\midrule
Prop. 1 & 2.7  & 2 & 35.0 & 0.33 \\

Prop. 2 & 5.3  & 4 & 33.5 & 0.33 \\

Prop. 3 & 7.1  & 6 & 31.1 & 0.33 \\

Prop. 4 & 8.3 & 8 & 29.5 & 0.33 \\

Prop. 5 & 9.4 & 10 & 28.6 & 0.33 \\

Prop. 6 & 10.0 & 15 & 26.3 & 0.33 \\
\bottomrule
\end{tabular}
\caption{Propagation parameters with $z_{\rm h}$ varying from $2\,{\rm kpc}$ to $15\,{\rm kpc}$ \cite{Fermi-LAT:2012pls, Huang:2016pxg}. Suppose a homogeneous spatial diffusion coefficient $D_{xx}=D_0\beta(E/4\,{\rm GeV})^\dt$, where $\beta$ is the Lorentz factor, $D_0$ is a coefficient, and $\dt=0.33$ reflects the Kolmogrov-type interstellar medium turbulence. The Alfvenic speed $v_{\rm A}$ characterizes the reacceleration effect. } \label{tab:1}
\end{table}

\begin{figure}
\centering \includegraphics[width=1\linewidth]{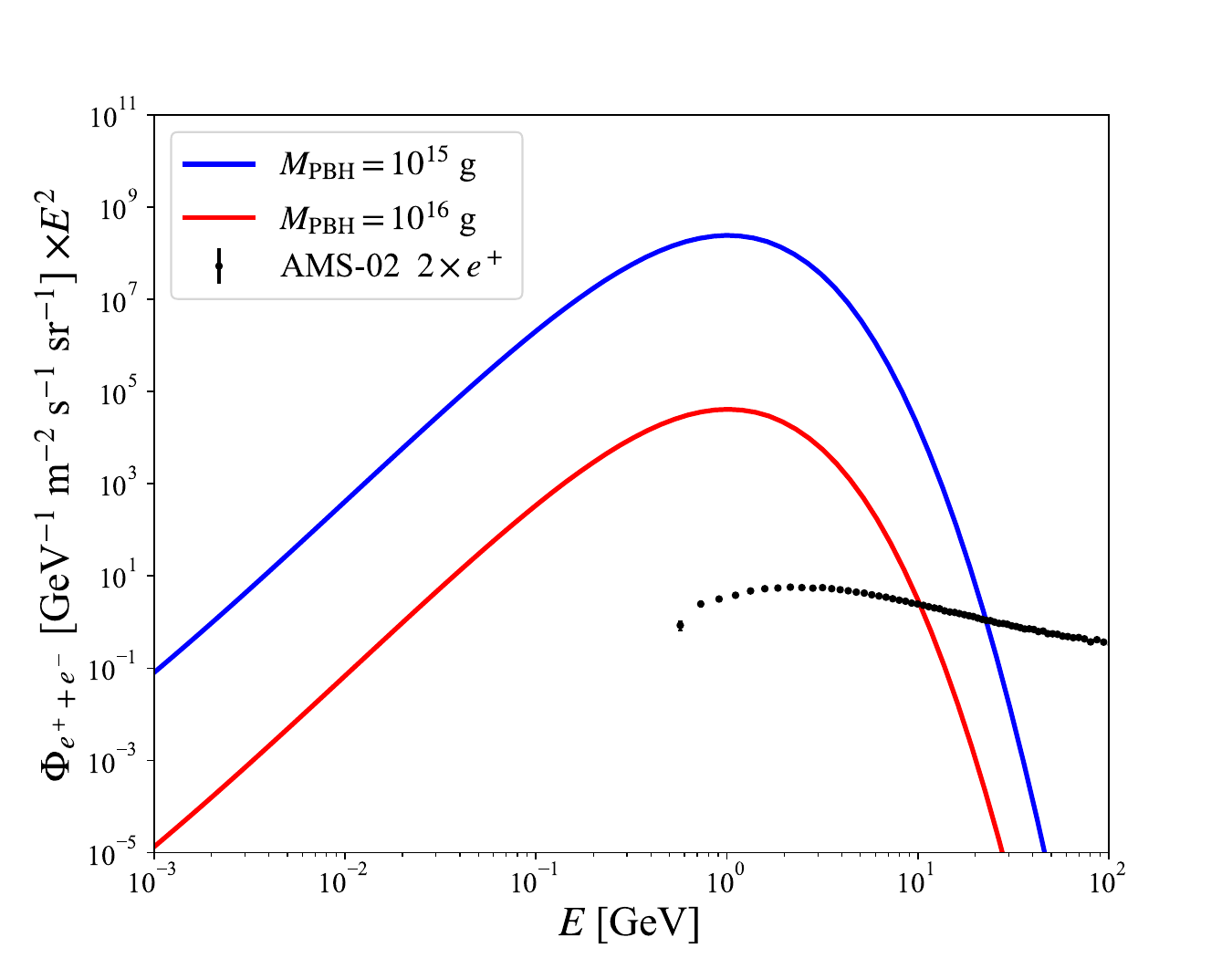}
\caption{The fluxes $\Phi_{e^++e^-}$ originating from the evaporation of PBHs with $M_\pb = 10^{15}\,{\rm g}$ and $10^{16}\,{\rm g}$, considering Prop. 6 in Tab. \ref{tab:1} alongside a solar modulation potential of $0.6\,{\rm GeV}$. The AMS-02 measurements are also presented for comparison \cite{AMS:2014xys}.} \label{fig:flux}
\end{figure}

Assuming all DM in the Milky Way consists of PBHs with a monochromatic mass distribution, we present the electron and positron spectra after propagation in Fig. \ref{fig:flux}. The fluxes $\Phi_{e^++e^-}$ are obtained with Prop. 6 in Tab. \ref{tab:1} and a solar modulation potential of $0.6\,{\rm GeV}$. Due to the reacceleration, the energies of the ejected electron/positron at sub-${\rm GeV}$ scale are boosted to the ${\rm GeV}$ range, aligning with the range covered by the AMS-02 data \cite{AMS:2013fma, AMS:2014gdf}. Therefore, the AMS-02 data can be employed to constrain the fraction of PBHs. As shown in Fig. \ref{fig:flux}, the flux $\Phi_{e^++e^-}$ decreases rapidly with the PBH mass $M_\pb$.

The astrophysical CR background encompasses conventional primary electrons, such as those originating from supernova remnants, as well as secondary electrons and positrons generated through inelastic collisions between CR nuclei and the interstellar medium. As we seek spectral features that stand out from the "smooth" background, it is reasonable to assume that the majority of the observational data can be well-fitted by the background. Following Ref. \cite{Huang:2016pxg}, we utilize the empirical model, which includes the primary electrons, secondary electrons/positrons, and the electron/positron excess from the extra source:
\begin{align}
\phi_{e^-}&=C_{e^-}E^{-\gamma_1^{e^-}}\lt[1+\lt(\f{E}{E_{\rm br}^{e^-}}\rt)^{\gamma_2^{e^-}}\rt]^{-1} ,\n\\
\phi_{e^+}&=C_{e^+}E^{-\gamma_1^{e^+}}\lt[1+\lt(\f{E}{E_{\rm br}^{e^+}}\rt)^{\gamma_2^{e^+}}\rt]^{-1} ,\n\\
\phi_{\rm s}&=C_{\rm s}E^{-\gamma^{\rm s}}\exp\lt(-\f{E}{E_{\rm c}^{\rm s}}\rt) .\n
\end{align}
Therefore, the total background energy spectrum of electrons plus positrons $\Phi_{{\rm bkg},\,e^\pm}$ is
\begin{align}
\Phi_{{\rm bkg},\,e^\pm}=\phi_{e^-}+1.6\phi_{e^+}+2\phi_{\rm s}.\n
\end{align}
where the factor 1.6 accounts for the asymmetry of electrons and positrons generated in $pp$ collisions \cite{Kamae:2006bf}. The best-fit parameters can be found in Tab. \uppercase\expandafter{\romannumeral3} of Ref. \cite{Huang:2016pxg}. 
In general, we directly fit the CR data with the model described above, focusing solely on the numerical shape without calculating the propagation of the background. In contrast, for the PBH source, we carefully calculate the propagation. When incorporating the contribution of the PBH source into the model, we optimize the fitting results by introducing the adjustment factors $\alpha_iE^{\bt_i}$, with $i=e^-$, $e^+$, and ${\rm s}$ correspond to $\phi_{e^-}$, $\phi_{e^+}$, and $\phi_{\rm s}$, respectively. This adjustment enables us to derive more conservative constraints.

\section{Results} \label{sec:res}

In this work, we constrain the PBH abundance $f_\pb$ through maximum likelihood fitting by utilizing data from the AMS-02 positron fraction \cite{AMS:2013fma} and total electron plus positron flux \cite{AMS:2014gdf}. We firstly calculate the $\chi^2$ with the inclusion of Hawking radiation contribution and then determine the best-fit value denoted as $\chi_0$. By requiring $\Delta\chi^2 =\chi^2-\chi_0^2>2.71$, we derive the 95\% confidence level upper limit on the PBH abundance. 

\begin{figure}
\centering
\subfigure[]{\includegraphics[width=1\linewidth]{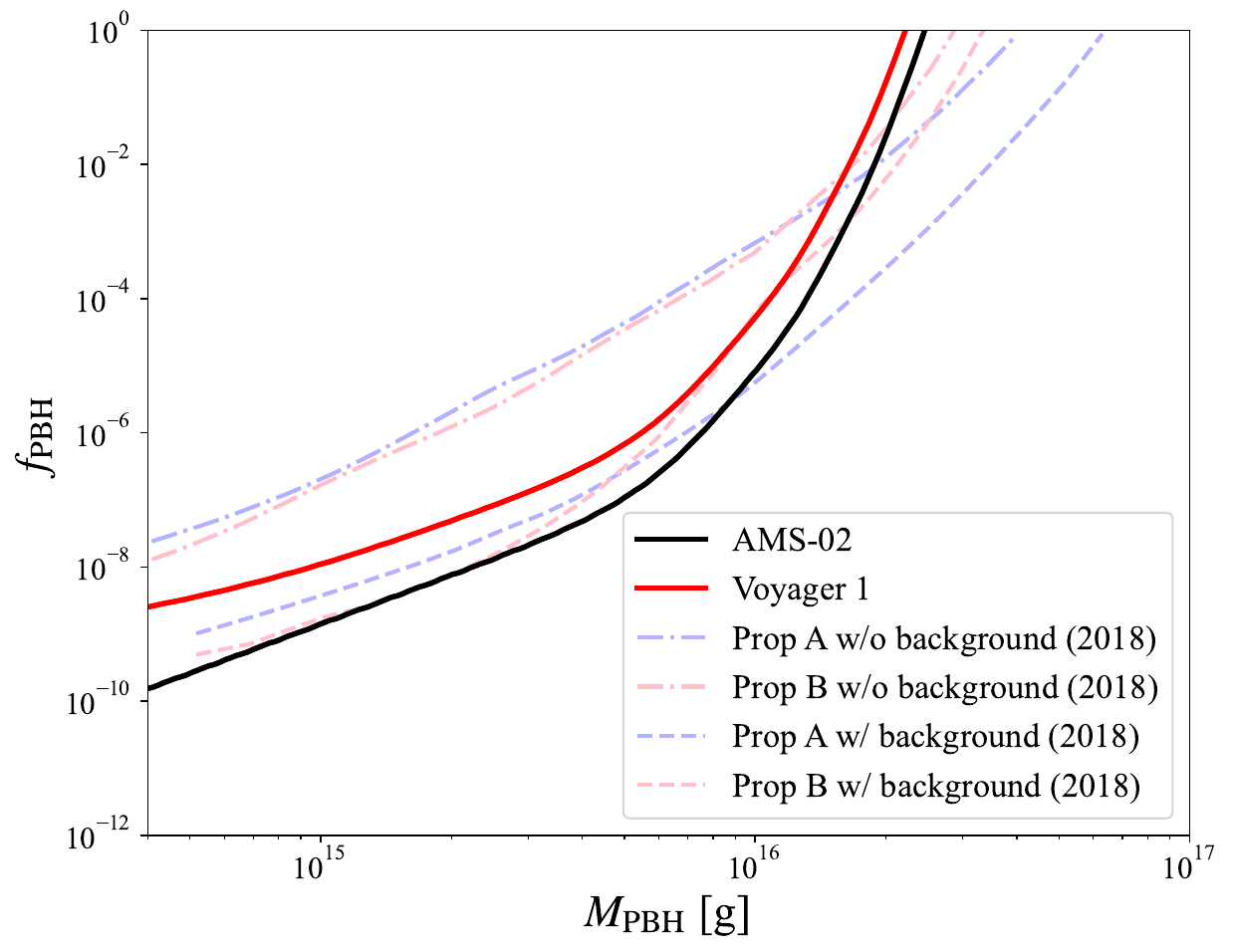} \label{fig:1}} \n\\
\subfigure[]{\includegraphics[width=1\linewidth]{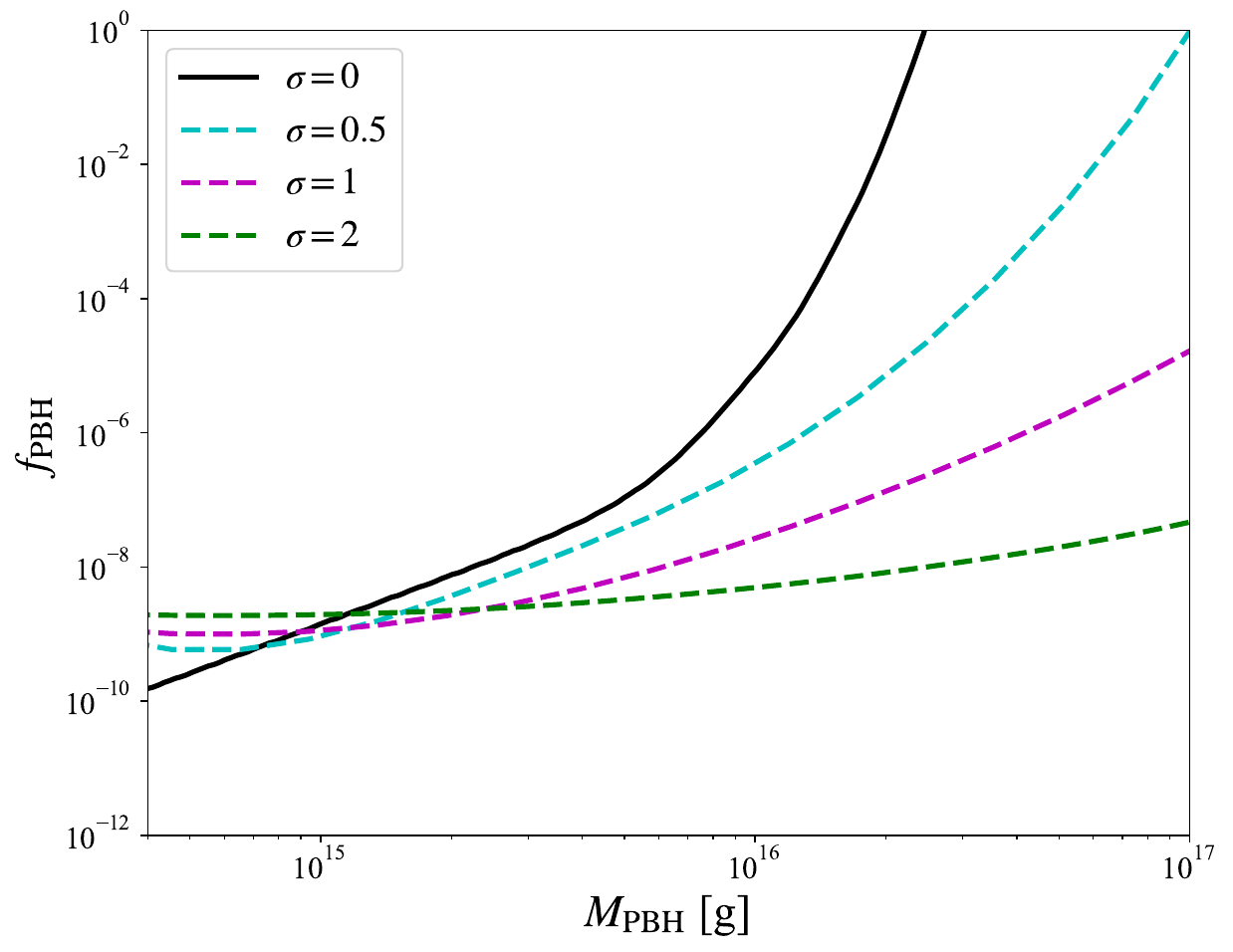} \label{fig:2}}
\caption{Constraints on PBHs abundance $f_\pb$ as a function of the PBH mass $M_\pb$, obtained in this study using AMS-02 data (solid black line) and Voyager 1 data (solid red line). The results from Ref. \cite{Boudaud:2018hqb} using Voyager 1 data are also incorporated (light blue and pink dashed lines with background, and dot-dashed lines without background). We adopt Prop. 6 in Tab. \ref{tab:1}, and set the solar modulation potential to $0.6\,{\rm GV}$ and $0\,{\rm GV}$ for the case with AMS-02 and Voyager 1 respectively. Fig. \ref{fig:1} assumes a monochromatic mass distribution, while Fig. \ref{fig:2} considers a log-normal mass distribution.} \label{fig:f}
\end{figure}

To begin with, we adopt Prop. 6 from Tab. \ref{tab:1} with a halo height $z_{\rm h}=15\,{\rm kpc}$ and set the solar modulation potential to $0.6\,{\rm GV}$. The constraint on $f_{\rm PBH}$ for monochromatic mass distribution of PBHs is depicted as the solid black line in Fig. \ref{fig:1}. For comparison, we also derive the limit from Voyager 1 data \cite{Stone:2013zlg, Cummings:2016pdr} with Prop. 6 (solid red line). Given that the Voyager 1 spacecraft has already crossed the heliopause threshold, the solar modulation potential is naturally set to $0\,{\rm GV}$. Additionally, constraints from Ref. \cite{Boudaud:2018hqb} based on the Voyager 1 data (light blue and pink lines) are incorporated. To maintain consistency with Ref. \cite{Boudaud:2018hqb}, we fit the Voyager 1 data with a power-law form energy spectrum as $\Phi_{{\rm Vbkg},\,e^\pm}=593.97E^{-1.31}$ (with $\chi_{\rm dof}^2=7.73/9$), omitting the adjustment factors. In contrast to the methodology employed in Ref. \cite{Boudaud:2018hqb}, our method refined by improving numerical calculation tools and incorporating the latest propagation parameters. Fig. \ref{fig:1} clearly illustrates the stronger constraint derived from the AMS-02 data compared to Voyager 1, particularly by more than an order of magnitude when $M_{\rm PBH}\lesssim10^{15}\,{\rm g}$. This superiority is largely attributed to the remarkably small errors in the AMS-02 measurements. Fig. \ref{fig:2} shows the constraints on $f_\pb$ for log-normal mass distribution PBHs portrayed by Eq. (\ref{log}). We set the central value of the log-normal distribution $\mu\lesssim 10^{17}\,{\rm g}$, and cut at $4\times 10^{14}\,{\rm g}$ as PBHs with lower masses have already evaporated. Various values of the width $\sg$ are taken into account, and the constraints notably strengthen with increasing $\sg$. For $\sg=0$, the constraint is consistent with that of monochromatic mass distribution (solid black lines in Figs. \ref{fig:1} and \ref{fig:2}).

\begin{figure}
\centering 
\subfigure[]{\includegraphics[width=1\linewidth]{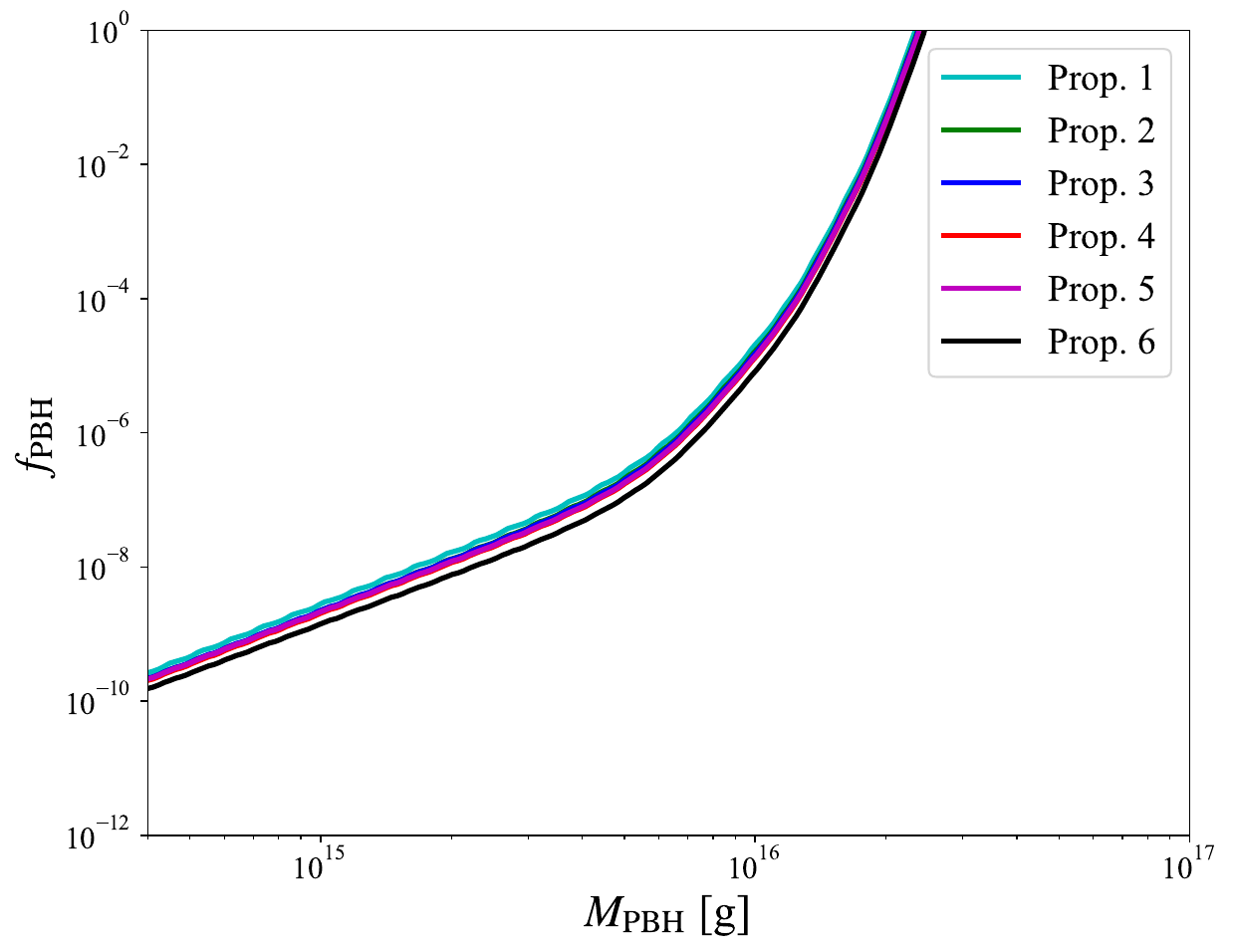} \label{fig:fpbhsix}} \quad
\subfigure[]{\includegraphics[width=1\linewidth]{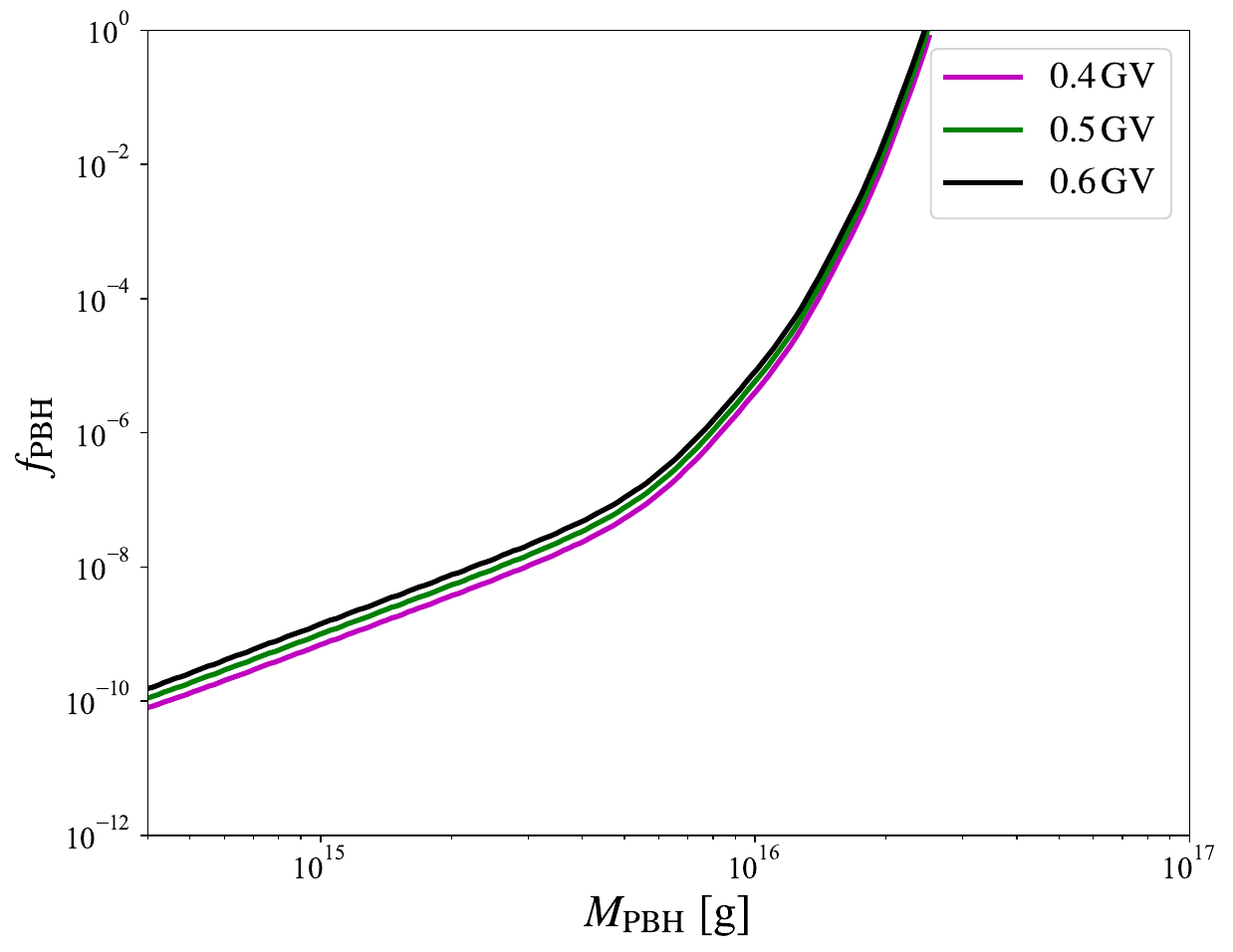} \label{fig:fpbhsun}}
\caption{Constraints on PBHs abundance $f_\pb$ as a function of the PBH mass $M_\pb$, with the monochromatic mass distribution assumption. Fig. \ref{fig:fpbhsix} adopts six groups of propagation parameters given in Tab. \ref{tab:1}, with a solar modulation potential of $0.6\,{\rm GV}$. Fig. \ref{fig:fpbhsun} utilizes Prop. 6 with three different solar modulation potentials.} \label{fig:fpbhs}
\end{figure}

For a more comprehensive analysis, with the assumption of monochromatic mass distribution, we investigate the constraints with different sets of propagation parameters and solar modulation potentials. The Props. 1 to 6 outlined in Tab. \ref{tab:1} are used to limit the PBH abundance, as illustrated in Fig. \ref{fig:fpbhsix}. Here, the solar modulation potential is fixed at $0.6\,{\rm GV}$. It can be seen that these different propagation parameters behave similarly when constraining the PBH abundance. Additionally, in accordance with Prop. 6, various cases of solar modulation are also considered in Fig. \ref{fig:fpbhsun}, with modulation potentials set at $0.4\,{\rm GV}$, $0.5\,{\rm GV}$, and $0.6\,{\rm GV}$, respectively. The limit on $f_\pb$ gradually weakens with modulation potential as more electrons and positrons are shielded by the solar magnetic field. Overall, different choices of propagation parameters or solar modulation within a reasonable range do not significantly impact the results as depicted in Figs. \ref{fig:fpbhs}.

\section{Conclusion} \label{sec:con}

PBHs with $M_\pb\lesssim 10^{16}\,{\rm g}$ are anticipated to inject sub-GeV electrons and positrons into the environment through Hawking radiation. Hypothesizing a diffusion plus reacceleration model of propagation, we have computed the fluxes for different PBH masses. The results reveal that part of sub-GeV electrons and positrons can be accelerated to the GeV level, falling within the observational range of AMS-02. In this work, we utilized the flux of these GeV electrons and positrons to constrain the PBH abundance $f_\pb$ in the Milky Way using AMS-02 data. Employing Prop. 6 in Tab. \ref{tab:1} and a solar modulation potential for $0.6\,{\rm GV}$, we have explored two cases involving monochromatic and log-normal mass distributions of PBHs. Our results limit $f_\pb\ll 0.1$ for $M_\pb \lesssim 10^{16}\,{\rm g}$, thus largely ruling out the possibility of PBHs within this mass range as a significant contributor to DM. We have also presented the constraints on $f_\pb$ with the remaining five sets of propagation parameters in Tab. \ref{tab:1}, demonstrating that different parameter choices do not significantly affect the results. Furthermore, while lower solar modulation potentials strengthen the limit on $f_\pb$, their impact is marginal within reasonable parameter space. 

\vspace{.5cm}
\noindent {\bf Acknowledgements} This work is supported by the National Key R\&D Program of China (Grant No. 2022YFF0503304), the National Natural Science Foundation of China (Grants No. 12373002, 12220101003, 11773075), the Youth Innovation Promotion Association of Chinese Academy of Sciences (Grant No. 2016288), and the Shandong Provincial Natural Science Foundation (Grant No. ZR2021MA021).

\vspace{.5cm}
\noindent {\bf Data Availability Statement} No Data associated in the manuscript.

\nocite{*}

\bibliographystyle{spphys}
\balance
\bibliography{ref}

\end{document}